\documentclass[journal]{journal}
\ifCLASSINFOpdf
   \usepackage[pdftex]{graphicx}
   \usepackage{pgfplots}
\else
\fi
\usepackage[cmex10]{amsmath}
\usepackage{amssymb}
\usepackage{algorithm}
\usepackage{algorithmic}
\usepgfplotslibrary{external}
\newcommand{\subjects}[0]{\ensuremath{\mathcal{S}}}
\newcommand{\subject}[0]{\ensuremath{s}}

\newcommand{\cells}[1]{\ensuremath{\mathcal{C}_{#1}}}
\newcommand{\cell}[0]{\ensuremath{c}}

\newcommand{\attr}[0]{\ensuremath{a}}
\newcommand{\attrname}[1]{\ensuremath{\texttt{#1}}}
\newcommand{\attrvalue}[1]{\ensuremath{\texttt{"#1"}}}

\newcommand{\features}[0]{\ensuremath{\mathcal{F}}}
\newcommand{\feature}[0]{\ensuremath{f}}

\newcommand{\histos}[0]{\ensuremath{\mathcal{H}}}
\newcommand{\histo}[1]{\ensuremath{H_{#1}}}

\newcommand{\distances}[0]{\ensuremath{{\bf D}}}
\newcommand{\distanceH}[0]{\ensuremath{d_\histos}}
\newcommand{\distanceXY}[0]{\ensuremath{d_\attrname{xy}}}
\newcommand{\distanceWass}[0]{\ensuremath{d_{wass}}}

\newcommand{\weight}[0]{\ensuremath{w}}

\newcommand{\prop}[0]{\ensuremath{P}}

\newcommand{\triangles}[1]{\ensuremath{\mathcal{T}_{#1}}}
\newcommand{\tr}[0]{\ensuremath{t}}

\newcommand{\graph}[1]{\ensuremath{\mathcal{G}_{#1}}}
\newcommand{\edges}[1]{\ensuremath{\mathcal{E}_{#1}}}

\newcommand{\values}[0]{\ensuremath{V}}
\newcommand{\val}[0]{\ensuremath{v}}

\newcommand{\tauAHC}[0]{\ensuremath{\tau_{\text{AHC}}}}
\newcommand{\tauDB}[0]{\ensuremath{\tau_{\text{DB}}}}

\begin{document}
%
\title{An Approach for Clustering Subjects According to Similarities in Cell Distributions within Biopsies}
%
%
%

\author{Yassine~El~Ouahidi\thanks{Y. El Ouahidi, M. Feller and B. Pasdeloup are with IMT Atlantique, Lab-STICC, UMR CNRS 6285, Brest F-29238, France. E-mail: yassine.el-ouahidi@imt-atlantique.net, matis.feller@imt-atlantique.net, bastien.pasdeloup@imt-atlantique.fr.},
        Matis~Feller,
        Matthieu~Talagas\thanks{M. Talagas is with the University of Brest, LIEN, F-29200 Brest, France; and the Department of Dermatology, Brest University Hospital, Brest, France. E-mail: matthieu.talagas@chu-brest.fr.},
        Bastien~Pasdeloup}

\maketitle
\thispagestyle{empty}

\begin{abstract}
In this paper, we introduce a novel and interpretable methodology to cluster subjects suffering from cancer, based on features extracted from their biopsies. Contrary to existing approaches, we propose here to capture complex patterns in the repartitions of their cells using histograms, and compare subjects on the basis of these repartitions. We describe here our complete workflow, including creation of the database, cells segmentation and phenotyping, computation of complex features, choice of a distance function between features, clustering between subjects using that distance, and survival analysis of obtained clusters. We illustrate our approach on a database of hematoxylin and eosin (H\&E)-stained tissues of subjects suffering from Stage I lung adenocarcinoma, where our results match existing knowledge in prognosis estimation with high confidence.
\end{abstract}


%
\IEEEpeerreviewmaketitle

\section{Introduction}
%
%
%
%
\IEEEPARstart{P}{ersonalized} oncololgy aims at improving the survival outcome of a subject, by understanding deeply their disease using attributes that are specific to their own metabolism. Among techniques that analyze biopsies of tumors, most works characterize subjects based on quantitative attributes, such as the number of tumor-infltrating lymphocytes (TILs) \cite{Mlecnik2011}, their density and area \cite{reichling2020artificial}, or the immunoscore \cite{galon2012immune}. However, it has been recently shown that more global organizations of cells -- and in particular TILs -- can have a strong impact on the survival prognosis \cite{Fridman2012, Yener2016, Saltz2018}.

In this work, we introduce a novel methodology to capture in details complex repartitions of cells within the tissues. Our approach can be used at various scales (cell level, clusters of cells level, etc.), thus providing numerous and complementary information regarding the tissue.

Contrary to existing approaches such as \cite{Saltz2018}, we do not classify the subjects in pre-defined classes. Instead, we propose to work in a non-supervised way, to group subjects based on their similarities regarding the defined features. This allows 1) exploration of new features that may correlate with survival; and 2) research of similar subjects for a newly seen person.

Our work is organized as follows: in Section \ref{sec:methodo}, we give an overview of our method and its inputs; in Section \ref{sec:features}, we define some features that capture particular organizations of cells; then, in Section \ref{sec:distances}, we explain which distance function we use, as well as our choice of clustering algorithm; finally, in Section \ref{sec:exp}, we illustrate our methodology on a database of H\&E-stained biopsies of subjects suffering from Stage I lung adenocarcinoma, and show that our results match existing knowledge in the field.

\section{Overview of the methodology}
\label{sec:methodo}

Biopsies are provided in the form of high-resolution scans of a tissue, which have been previously stained to reveal cell kernels. H\&E has been the most commonly used staining for years due to an easy and cheap acquisition of these images. Still, more complex techniques such as multiplex immunohistochemistry (mIHC) have been developed, which allow capturing richer information on the cells such as distinction of CD8/CD4/CD3 lymphocytes, albeit at a higher cost.

Once a high-resolution image has been acquired, a common practice is to perform its segmentation and phenotyping, in order to localize and characterze all cells. For H\&E images, tools using deep learning models \cite{Wang2019} have recently shown impressive performance in finding cells and associating them with a phenotype (cancer/stroma/lymphocyte). Proprietary tools exist for mIHC \cite{inForm}, as well as more recent solutions, here again involving deep learning models \cite{Mercadier2019}. All of these tools output similar information. For each cell kernel found, they provide at least the following attributes: \attrname{x} coordinate, \attrname{y} coordinate, and \attrname{phenotype}. Available phenotypes depend on the staining technique, and more attributes can be output depending on the tools (\emph{e.g.}, Keratin, PDL1, etc.).

Our methodology takes as an input a set $\subjects = \{ \subject^{(i)} \}_{i \in \{ 1, \dots, |\subjects|\}}$ of subjects. For each subject $\subject \in \subjects$, we have a set $\cells{\subject} = \{ \cell^{(i)} \}_{i \in \{ 1, \dots, |\cells{\subject}|\}}$ of cells. Each cell $\cell \in \cells{\subject}$ has a given number of attributes, which we can access as $\cell[\attr]$, with $\attr \in \{\attrname{x}, \attrname{y}, \attrname{phenotype}, \dots\}$.

From there, our method proceeds as follows:
\begin{enumerate}
    \item Let $\histos$ be the set of histograms. Design a set $\features = \{\feature^{(i)} : \subjects \rightarrow \histos \}_{i \in \{1, \dots, |\features| \}}$ of features of interest, which are functions that associate a subject $\subject \in \subjects$ with a histogram $\histo{\subject}^{(i)} \in \histos$. Examples of features are given in Section \ref{sec:features};
    \item Evaluate all features on all subjects to obtain histograms $\{ \histo{\subject}^{(i)} \}_{\subject \in \subjects, i \in \{1, \dots, |\features| \}}$.
    \item For each feature $\feature^{(i)} \in \features$, and each two distinct subjects $\subject^{(j)}, \subject^{(k)} \in \subjects$, compute the $|\subjects| \times |\subjects|$ distance matrix $\distances^{(i)}[j, k] = \distanceH\left(\histo{\subject^{(j)}}^{(i)}, \histo{\subject^{(k)}}^{(i)}\right)$. Details on the distance function $\distanceH$ are given in Section \ref{sec:distances};
    \item For each matrix $\distances^{(i)}$, feed it to a clustering algorithm, that will output a partition of $\subjects$ based on their similarity with respect to feature $\feature^{(i)}$. Details on the algorithm are given in Section \ref{sec:distances};
    \item Perform survival analysis of the sub-populations found this way to evaluate significance of the impact of $\feature^{(i)}$ on survival prognosis.
\end{enumerate}

Alternatively, step 3) can be changed to first compute a matrix $\distances$ which aggregates the various $\distances^{(i)}$ associated with individual features. This allows merging information captured by the features into one global matrix, which can be used as an input for the clustering algorithm. Details on the aggregation are given in Section \ref{sec:combination}.

\section{Proposed features}
\label{sec:features}

In this section, we propose five features that capture some interesting cell repartitions within the tissues\footnote{Codes for all subsequent functions will be made available upon acceptance.}. Each feature is a function $\feature^{(i)} : \subjects \rightarrow \histos$ that inputs a subject $\subject \in \subjects$ with cells $\cells{\subject}$, and outputs a histogram $\histo{\subject}^{(i)} \in \histos$. In the following, function $values\_to\_hist(\values)$ transforms a set $\values$ of values into a histogram which associates each value $\val \in \values$ with its number of occurrences in $\values$.

In order to simplify features definition, we first introduce a notion of filtering. Let $\cells{\subject}^- \subseteq \cells{\subject}$ be any subset of the cells $\cells{\subject}$ of subject $\subject \in \subjects$. We note $\cells{\subject}^-[\prop] \subseteq \cells{\subject}^-$ the subset of cells in $\cells{\subject}^-$ for which a proposition $\prop : \cells{\subject}^- \rightarrow \mathbb{B}$ is true.

As an example, remember from Section \ref{sec:methodo} that cells are given some attributes, including their phenotypes. Assume here the attribute \attrname{phenotype} can take values in $\{ \attrvalue{cancer}, \attrvalue{stroma}, \attrvalue{lymphocyte} \}$, we can define the following filter:
\begin{equation}
    \begin{array}[t]{lrcl}
        lymph : & \cells{\subject}^- & \rightarrow & \mathbb{B} \\
                    & \cell & \mapsto & \cell[\attrname{phenotype}] = \attrvalue{lymphocyte} \;.
    \end{array}
    \label{eq:lymph}
\end{equation}
Using this filter, we can get all lymphocytes in $\cells{\subject}$ as $\cells{\subject}[lymph]$. Similarly, we can easily define filters $stroma$ and $cancer$ that return cells of corresponding phenotypes.

\subsection{$\feature^{(1)}$: distances lymphocytes -- cancer cells}

Let us consider a subject $\subject \in \subjects$ with cells $\cells{\subject}$. In a first feature, we want to capture the proximity between lymphocytes and cancer cells. Each computed value will then be the minimum distance between a lymphocyte and all cancer cells.

To simplify this definition, we introduce a filter that, given a reference cell $\cell \in \cells{\subject}$, returns the cell within the set $\cells{\subject}^- \subseteq \cells{\subject}$ being filtered which minimizes Euclidean distance to $\cell$:
\begin{equation}
    \begin{array}[t]{lrcl}
        closest(\cell) : & \cells{\subject}^- & \rightarrow & \mathbb{B} \\
                    & \cell' & \mapsto & \cell' = \arg\min\limits_{\cell'' \in \cells{\subject}^-} \distanceXY(\cell'', \cell)
                    \;,
    \end{array}
    \label{eq:closest}
\end{equation}
where:
\begin{equation}
    \distanceXY(\cell, \cell') = \sqrt{(\cell'[\attrname{x}] - \cell[\attrname{x}])^2 + (\cell'[\attrname{y}] - \cell[\attrname{y}])^2} \;.
\end{equation}

Using this filter, Algorithm \ref{alg:1} details computation of $\histo{\subject}^{(1)}$. Informally, $\cells{\subject}[cancer][closest(\cell)]$ can be translated as \emph{the cancer cell which is the closest to the given cell $\cell$}.

\begin{algorithm}
\caption{$\feature^{(1)} (\subject)$}
\label{alg:1}
\begin{algorithmic}
    \STATE $\values := \{\}$
    \FORALL{$\cell \in \cells{\subject}[lymph]$}
        \STATE $d := \distanceXY(\cell,\cells{\subject}[cancer][closest(\cell)])$
        \STATE $\values := \values \cup \{ d \}$
    \ENDFOR
    \STATE $\histo{\subject}^{(1)} := values\_to\_hist(\values)$
    \RETURN $\histo{\subject}^{(1)}$
\end{algorithmic}
\end{algorithm}

\subsection{$\feature^{(2)}$: distances lymphocytes -- cancer/stroma interface}
\label{sec:feature_2}

In a second feature, we want to capture the proximity between lymphocytes and the cancer/stroma interface. Each computed value will then be the minimum distance between a lymphocyte and a cancer cell at that interface.

To simplify the definition of this feature we again introduce a new filter, that formalizes the notion of interface between cell types. Let $\cells{\subject}^- \subseteq \cells{\subject}$ and $\cells{\subject}^{-'} \subseteq \cells{\subject}$ be two non-intersecting subsets of the cells. Additionally, let $\triangles{\subject}$ be the set of triangles obtained by the Delaunay triangulation \cite{Delaunay1934} of the cells in $\cells{\subject}$, built from the \attrname{x} and \attrname{y} cell attributes (see Figure \ref{fig:rois} for an example of this triangulation). Each triangle $\tr \in \triangles{\subject}$ is the set of three cells that define it. We can define the filter:
\begin{equation}
    \begin{array}[t]{lrcl}
        inter(\cells{\subject}^{-'}) : & \cells{\subject}^- & \rightarrow & \mathbb{B} \\
                    & \cell & \mapsto & \exists \tr \in \triangles{\subject} : \cell \in \tr \\
                    & & & \text{and}~\exists \cell' \in \tr : \cell' \in \cells{\subject}^{-'}
                    \;,
    \end{array}
\end{equation}
which returns the subset of cells in $\cells{\subject}^-$ that are spatially close to cells in $\cells{\subject}^{-'}$, by checking existence of a triangle defined by cells of both subsets. This allows an easy definition of filters to obtain cells located at the border between two cell types, such as cancer and stroma for instance. Stroma cells at that interface are thus $\cells{\subject}[stroma][inter(\cells{\subject}[cancer])]$ and cancer ones $\cells{\subject}[cancer][inter(\cells{\subject}[stroma])]$.

Using this filter, Algorithm \ref{alg:2} details computation of $\histo{\subject}^{(2)}$. It is worth noting that we distinguish between lymphocytes that are within a cancer environment and those that are in a stroma environment by checking on which side of the interface they are located. The former will be associated with a negative distance in the histogram.

\begin{algorithm}
\caption{$\feature^{(2)} (\subject)$}
\label{alg:2}
\begin{algorithmic}
    \STATE $\values := \{\}$
    \FORALL{$\cell \in \cells{\subject}[lymph]$}
        \STATE $d_c := \distanceXY(\cell,\cells{\subject}[cancer][inter(\cells{\subject}[stroma])][closest(\cell)])$
        \STATE $d_s := \distanceXY(\cell,\cells{\subject}[stroma][inter(\cells{\subject}[cancer])][closest(\cell)])$
        \IF{$d_c < d_s$}
            \STATE $d_c := -d_c$
        \ENDIF
        \STATE $\values := \values \cup \{ d_c \}$
    \ENDFOR
    \STATE $\histo{\subject}^{(2)} := values\_to\_hist(\values)$
    \RETURN $\histo{\subject}^{(2)}$
\end{algorithmic}
\end{algorithm}

\subsection{$\feature^{(3)}$: distances between aggregates of lymphocytes}

This third feature catpures the proximity between dense aggregates of cells. Each computed value will be the minimum distance between an aggregate of lymphocytes and the other aggregates of such cells.

Once again, we simplify the definition of the feature by introducing a filter that will only keep cells that belong to the same aggregate. To do so, we again use the Delaunay triangulation of the cells. Let $\edges{\subject}$ be the set of edges in the Delaunay triangulation. Then, let $\cells{\subject}^- \subseteq \cells{\subject}$ be a subset of the cells, and let $\edges{\subject}^- \subseteq \edges{\subject}$ be the subset of edges that connect two cells in $\cells{\subject}^-$. The graph $\graph{\subject}$ made of vertices $\cells{\subject}^-$ and edges $\edges{\subject}^-$ consists of disjoint connected components, which we note $\{ \graph{\subject}^{(i)} \}_{i \in \{1, \dots, |\graph{\subject}|\}}$. We now define the following filter:
\begin{equation}
    \begin{array}[t]{lrcl}
        cc(i) : & \cells{\subject}^- & \rightarrow & \mathbb{B} \\
                    & \cell & \mapsto & \cell \in \graph{\subject}^{(i)}
                    \;.
    \end{array}
\end{equation}

Using this filter, Algorithm \ref{alg:3} details computation of $\histo{\subject}^{(3)}$. Informally, $\cells{\subject}[lymph][cc(j)][closest(\cell)]$ can be translated as \emph{the cell in the $j^{th}$ aggregate of lymphocytes which is the closest to the given cell \cell}.

\begin{algorithm}
\caption{$\feature^{(3)} (\subject)$}
\label{alg:3}
\begin{algorithmic}
    \STATE $\values := \{\}$
    \FORALL{$i$}
        \STATE $d_i := \infty$
        \FORALL{$j$, $j \neq i$}
            \STATE $d_{ij} := \infty$
            \FORALL{$\cell \in \cells{\subject}[lymph][cc(i)]$}
                \STATE $d_{ij} := \min(d_{ij},  \distanceXY(\cell, \cells{\subject}[lymph][cc(j)][closest(\cell)]))$
            \ENDFOR
            \STATE $d_i := \min(d_i, d_{ij})$
        \ENDFOR
        \STATE $\values := \values \cup \{ d_i \}$
    \ENDFOR
    \STATE $\histo{\subject}^{(3)} := values\_to\_hist(\values)$
    \RETURN $\histo{\subject}^{(3)}$
\end{algorithmic}
\end{algorithm}

\subsection{$\feature^{(4)}$: sizes of the aggregates of lymphocytes}

This fourth feature catpures the variety of sizes among aggregates of lymphocytes. Each computed value will be the number of lymphocytes in an aggregate of such cells.

Algorithm \ref{alg:4} details computation of $\histo{\subject}^{(4)}$.

\begin{algorithm}
\caption{$\feature^{(4)} (\subject)$}
\label{alg:4}
\begin{algorithmic}
    \STATE $\values := \{\}$
    \FORALL{$i$}
        \STATE $n_i := |\cells{\subject}[lymph][cc(i)]|$
        \STATE $\values := \values \cup \{ n_i \}$
    \ENDFOR
    \STATE $\histo{\subject}^{(4)} := values\_to\_hist(\values)$
    \RETURN $\histo{\subject}^{(4)}$
\end{algorithmic}
\end{algorithm}

\subsection{$\feature^{(5)}$: densities of lymphocytes at cancer/stroma interface}

Finally, this fifth feature catpures the densities of lymphocytes in bands around the cancer/stroma interface. Each computed value will be the density of lymphocytes among all cells located in a certain interval of distance from interface cells. As for feature $\feature^{(2)}$, we make the distinction between bands within the cancer environment and those within the stroma environment, by associating negative bins with the former. We have chosen to consider bands of $20\mu m$ to include multiple cells (as kernel size of the considered tissues is approximately $7-8 \mu m$).

Algorithm \ref{alg:5} details computation of $\histo{\subject}^{(5)}$.

\begin{algorithm}
\caption{$\feature^{(5)} (\subject)$}
\label{alg:5}
\begin{algorithmic}
    \STATE $band\_width := 20$
    \STATE $nb\_cells := \{\}$
    \STATE $nb\_lymph := \{\}$
    \FORALL{$\cell \in \cells{\subject}$}
        \STATE $d_c := \distanceXY(\cell,\cells{\subject}[cancer][inter(\cells{\subject}[stroma])][closest(\cell)])$
        \STATE $d_s := \distanceXY(\cell,\cells{\subject}[stroma][inter(\cells{\subject}[cancer])][closest(\cell)])$
        \IF{$d_c < d_s$}
            \STATE $d_c := -d_c$
        \ENDIF
        \STATE $band := d_c // band\_width$~~~~\# Euclidean division
        \IF{$band \not\in nb\_cells$}
            \STATE $nb\_cells[band] := 0$
            \STATE $nb\_lymph[band] := 0$
        \ENDIF
        \STATE $nb\_cells[band] := nb\_cells[band] + 1$
        \IF{$\cell[\attrname{phenotype}] = \attrvalue{lymphocyte}$}
            \STATE $nb\_lymph[band] := nb\_lymph[band] + 1$
        \ENDIF
        \STATE $\histo{\subject}^{(5)} := \{\}$
        \FORALL{$band \in nb\_cells$}
            \STATE $\histo{\subject}^{(5)}[band] := \frac{nb\_lymph[band]}{nb\_cells[band]}$
        \ENDFOR
    \ENDFOR
    \RETURN $\histo{\subject}^{(5)}$
\end{algorithmic}
\end{algorithm}

\section{Distances between features, and clustering}
\label{sec:distances}

\subsection{A distance function between histograms}

In Section \ref{sec:features}, we have introduced a few features that capture various repartitions of particular types of cells. Each of these features takes the form of a histogram, which is a richer representation than simple numerical features.

The idea of what comes next is to cluster subjects based on the repartitions of cells captured by these histograms. This implies the necessity to choose a distance function $\distanceH$ between histograms that complies with the captured information, \emph{i.e.}, for two histograms $\histo{\subject}^{(i)}$ and $\histo{\subject'}^{(i)}$ -- obtained by evaluating feature $\feature^{(i)}$ on two subjects $\subject, \subject' \in \subjects$ -- we want the distance $\distanceH(\histo{\subject}^{(i)}, \histo{\subject'}^{(i)})$ to be low when both histograms have similar shapes around the same values; and to increase as their shapes or locations differ.

A good candidate is therefore the Wasserstein distance between distributions (see \emph{e.g.}, \cite{Peyre2019}). This distance measures the quantity of work required to transform a distribution into another one. More formally, given two (positive, unit-sum) histograms $\histo{1}$ and $\histo{2}$, it is obtained by solving an optimization problem defined as:
\begin{equation}
    \begin{aligned}
    \distanceWass(\histo{1}, \histo{2}) = \min_\textbf{T} \sum_{h_1, h_2} \textbf{T}[h_1, h_2] \textbf{C}[h_1, h_2]\\s.t.~~\textbf{T} \textbf{1} = \histo{1};~~\textbf{T}^\top \textbf{1} = \histo{2};~~\textbf{T}\geq 0 \;,
    \end{aligned}
    \label{eq:wasserstein}
\end{equation}
where $\textbf{C}$ is the matrix that defines the cost to move a unit of mass from the $h_1^{th}$ bin of distribution $\histo{1}$ to the $h_2^{th}$ bin of distribution $\histo{2}$. A solution to the problem is a matrix $\textbf{T}$ of optimal transport between histogram $\histo{1}$ and $\histo{2}$. This problem is of complexity $\mathcal{O}(n^3)$. However, solvers are very efficient and make its resolution very tractable \cite{Flamary2017}.

As this distance is defined on distributions, we must therefore normalize our histograms so that the sum of values equals $1$. A direct consequence of this is that we cannot distinguish some histograms anymore. A simple example is -- on one side -- a histogram consisting of a unique bin at $0$ with value $10$; and -- on the other side -- a histogram consisting of a unique bin at $0$ with value $100$. After normalization, both distributions feature a unique bin at $0$ of value $1$.

At first glance, this may seem problematic, as some features such as $\feature^{(1)}$ enforce a direct relation between the number of lymphocytes in the tissue and the number of histogram entries. However, such quantitative aspects can be included as numerical features during classification if one wants to include the total number of lymphocytes as a feature. Additionally, this normalization does not change the shape of the histogram, which allows us to compare -- taking again the example of $\feature^{(1)}$ -- the repartitions of lymphocytes around cancer cells.

In the remaining of this document, we will therefore use Wasserstein distance $\distanceWass$ between our histograms for $\distanceH$, assuming both histograms $\histo{\subject}^{(i)}$ and $\histo{\subject'}^{(i)}$ -- obtained by evaluating feature $\feature^{(i)}$ on two subjects $\subject, \subject' \in \subjects$ -- have previously been normalized, \emph{i.e.},
\begin{equation}
    \distanceH\left(\histo{\subject}^{(i)}, \histo{\subject'}^{(i)}\right) = \distanceWass\left(\frac{\histo{\subject}^{(i)}}{\left|\histo{\subject}^{(i)}\right|_1}, \frac{\histo{\subject'}^{(i)}}{\left|\histo{\subject'}^{(i)}\right|_1}\right)\;,
\end{equation}
where $|\cdot|_1$ denotes the $\ell_1$ norm of all values in the histogram.

The cost matrix \textbf{C} in Equation \ref{eq:wasserstein} is chosen to be linear, \emph{i.e.} if there is a bin at value $h_1$ and a bin at value $h_2$, the cost $\textbf{C}[h_1, h_2]$ for transporting a unit of mass between $h_1$ and $h_2$ is $| h_1 - h_2 |$. Dimension of $\textbf{C}$ is chosen so that each distinct value in $\histo{\subject}^{(i)}$ and $\histo{\subject'}^{(i)}$ has its own bin.

\subsection{Clustering of the subjects}

Now that we have defined a distance function that compares the shapes of distributions captured by our features, we can cluster all subjects in $\subjects$ -- in a non-supervised way -- to group together those that have similar repartitions of cells.

For each feature $\feature^{(i)} \in \features$, and any pair of subjects $\subject^{(j)}, \subject^{(k)} \in \subjects$, we can now compute a $|\subjects| \times |\subjects|$ matrix:
\begin{equation}
    \distances^{(i)}[j, k] = \distanceH\left(\histo{\subject^{(j)}}^{(i)}, \histo{\subject^{(k)}}^{(i)}\right) \;.
\end{equation}

We can now feed $\distances^{(i)}$ to a clustering algorithm to assess the abiliy of the feature to create meaningful sub-populations. Because of its high interpretability, we have chosen to use ascending hierarchical classification (AHC) with single-linkage \cite{Szekely2005} as our clustering algorithm. Starting with each subject into their own cluster, this algorithms iteratively merges clusters -- chosen to minimize dissimilarity between merged clusters -- until the entire population belongs to the same cluster. This way, it produces a dendrogram, in which the leaves are individual subjects. As a consequence, the deeper we go in the dendrogram (toward its leaves), the more homogeneous the clusters we find are.

Based on $\distances^{(i)}$, we use AHC to partition $\subjects$ into two disjoint sub-populations $\subjects = \subjects^{(i)}_1 \sqcup \subjects^{(i)}_2$ as follows:
\begin{itemize}
    \item The first sub-population $\subjects^{(i)}_1 \subseteq \subjects$ is obtained by cutting the dendrogram at a threshold $\tauAHC$, and keeping the deepest resulting sub-tree. By construction, this sub-population will be homogeneous with respect to the feature used to build the input distances matrix;
    \item The second sub-population $\subjects^{(i)}_2 \subseteq \subjects$ is obtained by merging all subjects that do not belong to the first one. By construction, these subjects will be dissimilar with respect to the feature.
\end{itemize}

The reasoning behind this approach is that we want to verify whether having one particular distribution of cells is characteristic of the survival prognosis, assuming the subjects without that particular distribution may have very personal -- diverse -- profiles.

In order to determine the value of the threshold $\tauAHC$ that will separate sub-populations, we have made an exhaustive search over its possible values. Then, for each two resulting sub-populations $\subjects^{(i)}_1$ and $\subjects^{(i)}_2$, we have computed their Kaplan-Meier survival curves \cite{Kaplan1958}, and have performed a log-rank test \cite{Mantel1966} to assess how different they are. We have then kept the value of $\tauAHC$ which minimizes the $p$-value of that test, \emph{i.e.}, which most significantly separates the two survival curves. The idea behind this choice is to maximize the homogeneity of $\subjects^{(i)}_1$ without any prior on the size of the clusters.

\subsection{Using multiple features}
\label{sec:combination}

In the previous paragraphs, we have explained how to cluster subjects based on a single matrix of distances, associated with one particular feature. In practice, we would like to combine the various features we have defined, to group together subjects that tend to behave similarly on multiple aspects.

To do so, we want to build a matrix $\distances$ that captures the fact that two subjects tend to belong to the same clusters $\subjects^{(i)}_1$ or $\subjects^{(i)}_2$, for each feature $\feature^{(i)} \in \features$. Additionally, we want to be able to give more importance to some features in that global matrix, to allow a pathologist to inject additional knowledge in the aggregation.

Let $\subject^{(j)}, \subject^{(k)} \in \subjects$ be two subjects. The requirements described previously can be integrated in the computation of $\distances$ as follows:
\begin{equation}
    \distances[j, k] = \sum\limits_{i = 1}^{|\features|} \weight^{(i)} \cdot same\left(\subject_j, \subject_k, \subjects^{(i)}_1\right) \;,
    \label{eq:global}
\end{equation}
where:
\begin{equation}
    same\left(\subject_j, \subject_k, \subjects^{(i)}_1\right) =
    \left\{
    \begin{array}{ll}
          0 & \text{if}~(\subject_j \in \subjects^{(i)}_1~\text{and}~\subject_k \in \subjects^{(i)}_1) \\
          & \text{or}~(\subject_j \not\in \subjects^{(i)}_1~\text{and}~\subject_k \not\in \subjects^{(i)}_1) \\
          1 & \text{otherwise} \;, \\
    \end{array}
    \right.
\end{equation}
and where $\{ \weight^{(i)}\}_{i \in \{1, \dots, |\features|}$ are given real values, weighting the importance of features in the combination. Their values can be determined either using expert knowledge, or more agnostic methods, as proposed in Section \ref{sec:combination_results}.

In Equation \ref{eq:global}, the higher the value in $\distances[j, k]$, the more subjects $\subject^{(j)}$ and $\subject^{(k)}$ tend to belong to different clusters.

\section{Experiments}
\label{sec:exp}

\subsection{Database used}

In order to evaluate our approach, we have chosen to focus on Stage I lung adenocarcinoma. We have extracted all corresponding subjects from the Cancer Genome Atlas (TCGA) repository \cite{TCGA} to create a population $\subjects$ of $140$ subjects. For each of these subjects, the TCGA repository provides us with high resolution scans of the subject's biopsy, stained with H\&E, as well as clinical information, including the subject's time to last follow-up (TTLFU) and alive state at that time. When combined, these two pieces of information allow us to compute the Kaplan-Meier survival estimator of the population.

From these $140$ subjects, we have chosen to remove those that were alive with a too small TTLFU (for which we don't have enough hindsight), as well as those that were dead with a too high TTLFU (as their tumor may have evolved too much since diagnosis). We have therefore set up a threshold TTLFU $\tauDB$, below which alive subjects are dropped, and above which dead subjects are dropped. The value of $\tauDB = 366$ days has been chosen to maximize the database size while encouraging these criteria. After this filtering, our database consists of $115$ subjects, $97$ of which have an alive state.

\begin{figure*}[!t]
    \centering
    \includegraphics[height=0.37\textwidth]{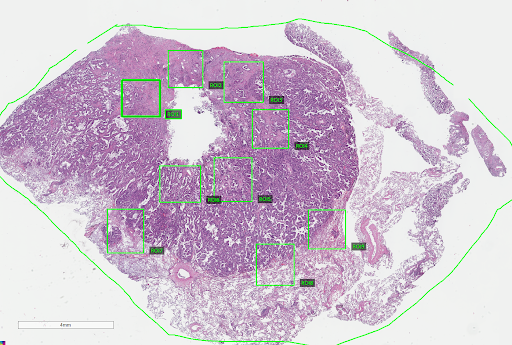}
    \includegraphics[height=0.37\textwidth]{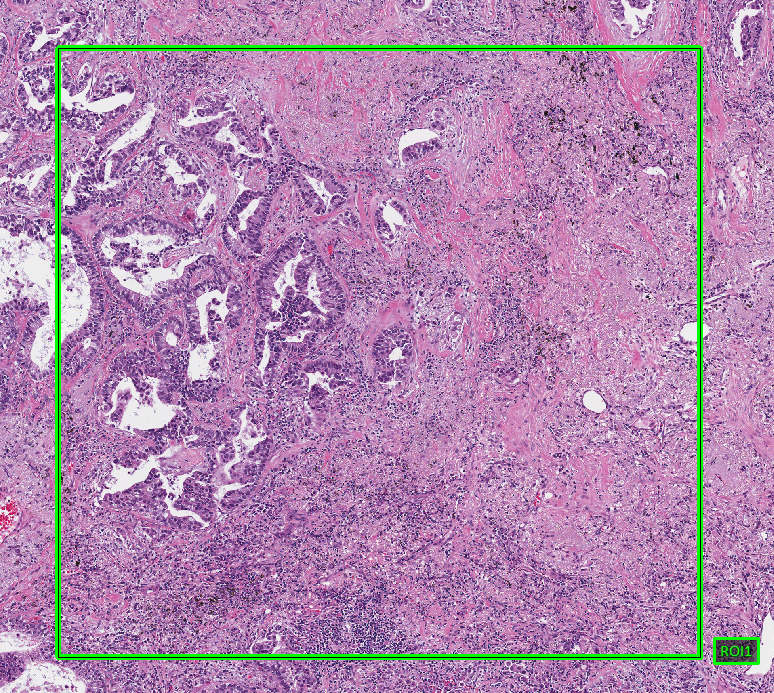} ~~ \\
    \includegraphics[width=\textwidth]{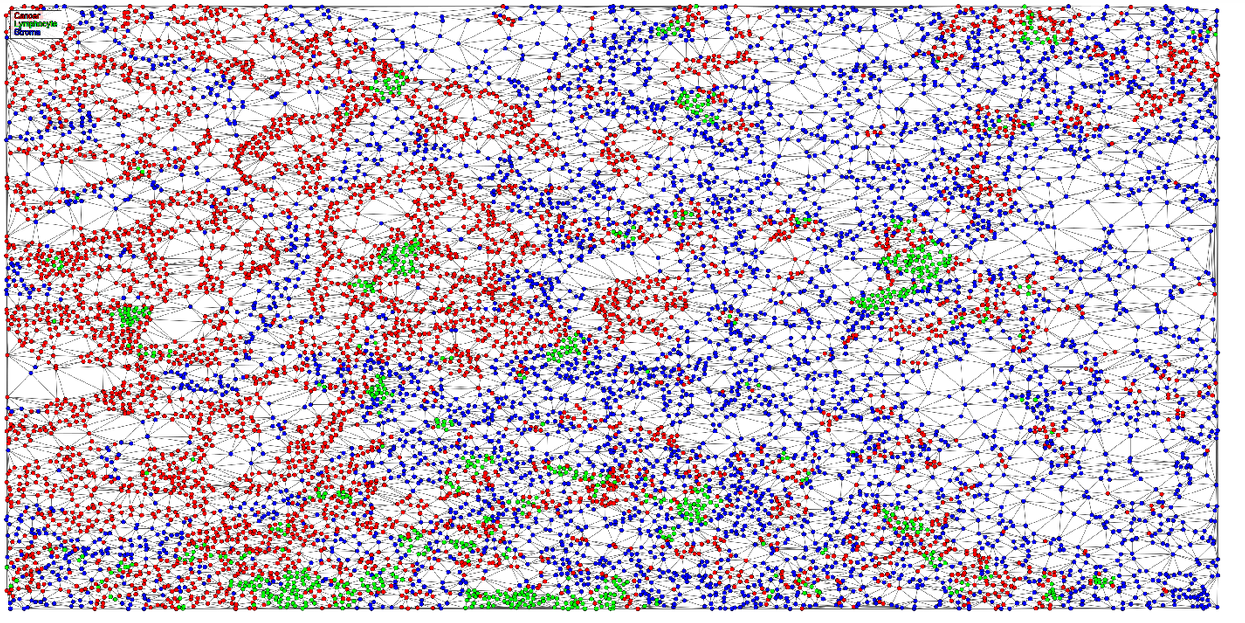}
    \caption{Manual selection of ROIs on a subject's SVS slide (top left), closeup on one of the selected ROIs (top right), and representation of the Delaunay triangulation of the cells in that ROI (bottom). In that representation, lymphocytes are depicted in green, cancer cells in red and stroma cells in blue. ROIs are selected to include both cancer and stroma cells in reasonable quantities.}
    \label{fig:rois}
\end{figure*}

For each subject, we have considered one scan of their biopsy, and have defined multiple regions of interest (ROIs), chosen at hand to be located in areas that appear to include both cancer and stroma cells, mostly located at the tumor border. The number of ROIs defined this way depends on the size and contents of the image. Figure \ref{fig:rois} depicts an example of ROI selection in a tissue.

For each identified ROI, we have segmented and phenotyped the cells in presence using the ConvPath tool \cite{Wang2019}. Each cell is therefore associated with the following attributes: \attrname{x} coordinate, \attrname{y} coordinate and \attrname{phenotype}, that last one taking values in $\{ \attrvalue{cancer}, \attrvalue{stroma}, \attrvalue{lymphocyte} \}$.

Finally, to enforce our prior during ROIs selection, we have filtered out those in which the ratio of cancer/stroma cells was not in the $[0.3, 0.7]$ interval. After this second filtering, we have as an input a database of $94$ subjects, $80$ of which have an alive state. Each of them is associated with $1$ to $9$ ROIs, with an average of $3.86$ ROI per subject. In the remaining of these documents, features for a subject will be computed per ROI, and aggregated in a single histogram before normalization.

\subsection{Individual results per feature}

First, we want to assess if the features we have proposed in Section \ref{sec:features} are significantly correlating with survival prognosis. To do so, we apply our methodology for each feature proposed in Section \ref{sec:features}.

Table \ref{tab:results} presents the obtained results per feature. For each feature $\feature^{(i)} \in \features$, we report respective sizes of resulting sub-populations $\subjects^{(i)}_1$ and $\subjects^{(i)}_2$. Significance of the clustering is analyzed through the log-rank test of Kaplan-Meier estimators between both sub-populations. The $p$-value of that test is given in the last column\footnote{All codes used to obtain these results will be made available upon acceptance.}.

\begin{table}[!t]
    \renewcommand{\arraystretch}{1.6}
    \caption{Survival analysis of the population, based on the proposed features, taken individually.}
    \label{tab:results}
    \centering
    \begin{tabular}{|c|c|c|c|}
        \hline
        Feature & $\frac{\left|\subjects^{(i)}_1\right|}{|\subjects|}$ & $\frac{\left|\subjects^{(i)}_2\right|}{|\subjects|}$ & $p$-value \\
        \hline
        $\feature^{(1)}$ & $0.55$ & $0.45$ & $0.0574$ \\
        \hline
        $\feature^{(2)}$ & $0.31$ & $0.69$ & $0.0002$ \\
        \hline
        $\feature^{(3)}$ & $0.52$ & $0.48$ & $0.0124$ \\
        \hline
        $\feature^{(4)}$ & $0.64$ & $0.36$ & $0.0151$ \\
        \hline
        $\feature^{(5)}$ & $0.19$ & $0.81$ & $0.0005$ \\
        \hline
    \end{tabular}
\end{table}

Results show that features $\feature^{(2)}$, $\feature^{(3)}$, $\feature^{(4)}$ and $\feature^{(5)}$ separate survival curves significantly ($p$-value $< 0.05$). More precisely, $\feature^{(2)}$ and $\feature^{(5)}$ have a strong statistical significance ($p$-value $\ll 0.05$).

\begin{figure*}[!t]
    \centering
    \input{survival_f1.pgf} \\
    \scalebox{0.47}{\input{f1_cluster1.pgf}} \hfill \includegraphics[height=5.4cm]{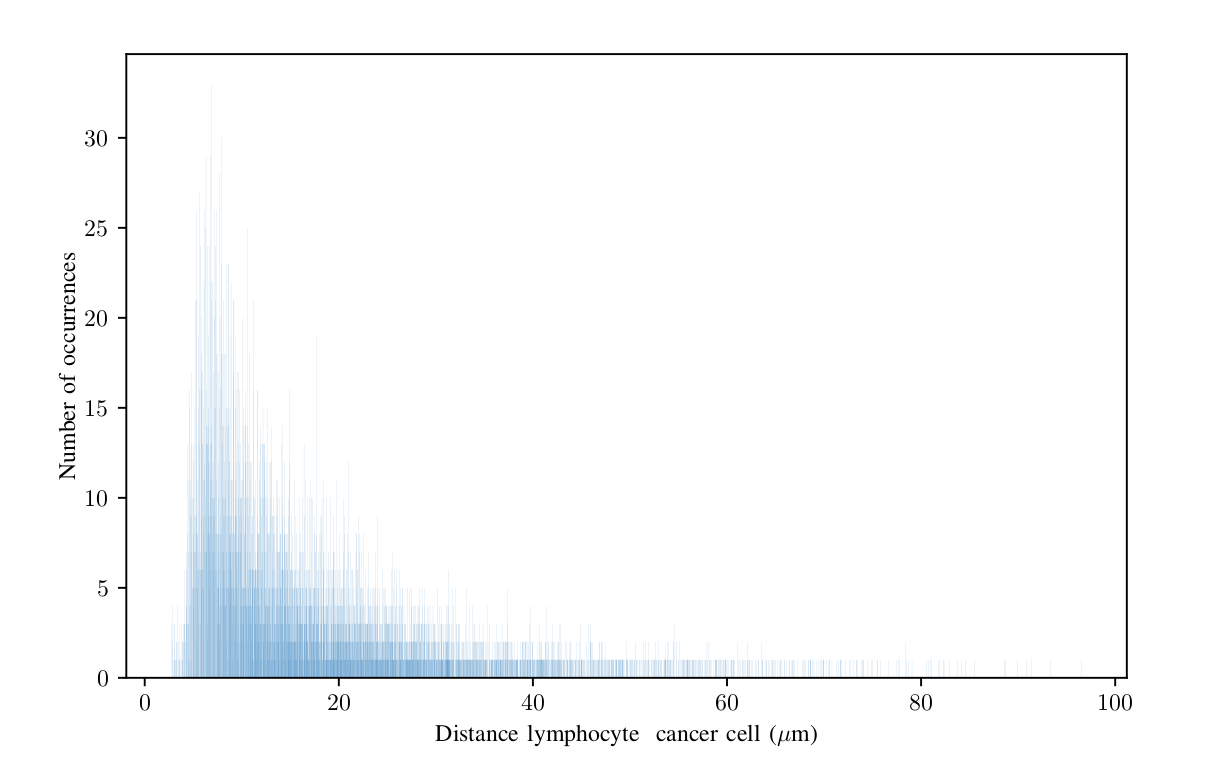}
    \caption{Survival curves associated with both sub-populations $\subjects^{(1)}_1$ and $\subjects^{(1)}_2$ obtained by AHC using $\distances^{(1)}$ (top); histogram $\histo{\subject_1^*}^{(1)}$ associated with the most central subject $\subject_1^* \in \subjects^{(1)}_1$ (alive, TTLFU $2973$ days) of the first sub-population $\subjects^{(1)}_1$, obtained by feeding $\distances^{(1)}$ to the AHC clustering algorithm (bottom left); and corresponding histogram $\histo{\subject_2^*}^{(1)}$ (alive, TTLFU $466$ days) of the central subject $\subject_2^* \in \subjects^{(1)}_2$ of the other sub-population $\subjects^{(1)}_2$ (bottom right).}
    \label{fig:histo_f1}
\end{figure*}

\begin{figure*}[!t]
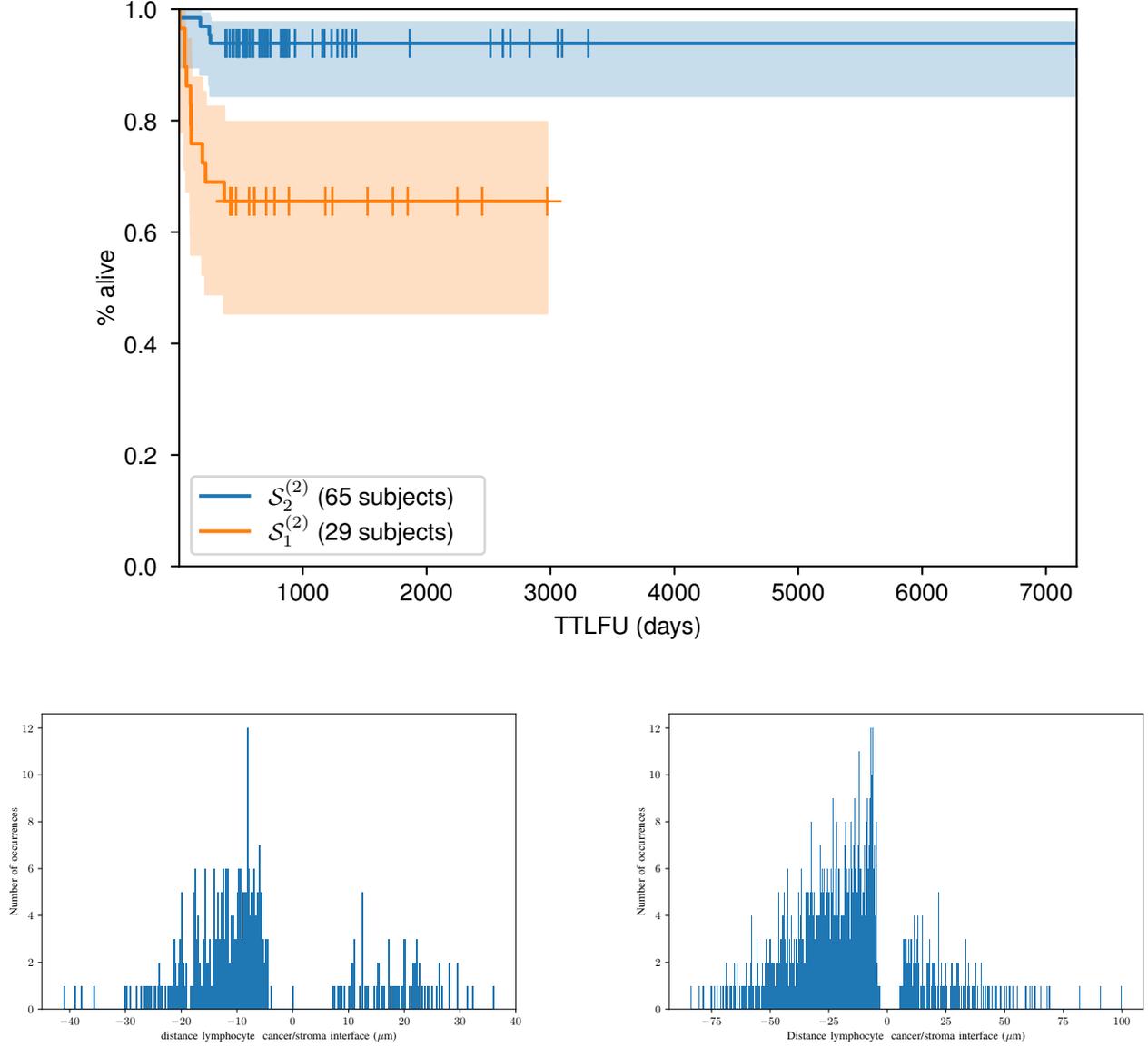

    \centering
    \input{survival_f2.pgf} \\
    \scalebox{0.47}{\input{f2_cluster1.pgf}} \hfill \scalebox{0.47}{\input{f2_cluster2.pgf}}
    \caption{Survival curves associated with both sub-populations $\subjects^{(2)}_1$ and $\subjects^{(2)}_2$ obtained by AHC using $\distances^{(2)}$ (top); histogram $\histo{\subject_1^*}^{(2)}$ associated with the most central subject $\subject_1^* \in \subjects^{(2)}_1$ (alive, TTLFU $2973$ days) of the first sub-population $\subjects^{(2)}_1$, obtained by feeding $\distances^{(2)}$ to the AHC clustering algorithm (left); and corresponding histogram $\histo{\subject_2^*}^{(2)}$ of the central subject $\subject_2^* \in \subjects^{(2)}_2$ (alive, TTLFU $670$ days) of the other sub-population $\subjects^{(2)}_2$ (right).}
    \label{fig:histo_f2}
\end{figure*}

\begin{figure*}[!t]
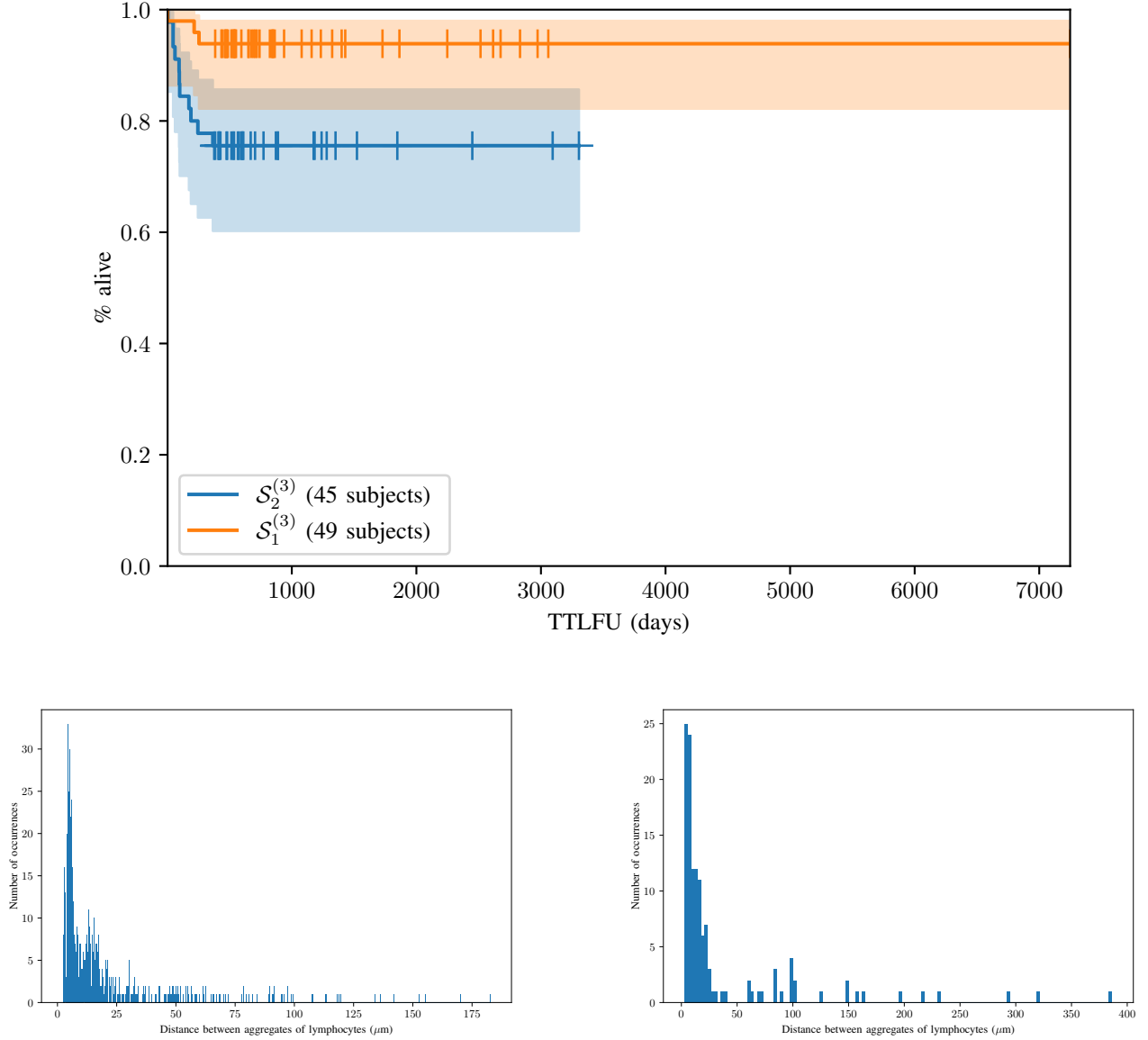

    \centering
    \input{survival_f3.pgf} \\
    \scalebox{0.47}{\input{f3_cluster1.pgf}} \hfill \scalebox{0.47}{\input{f3_cluster2.pgf}}
   \caption{Survival curves associated with both sub-populations $\subjects^{(3)}_1$ and $\subjects^{(3)}_2$ obtained by AHC using $\distances^{(3)}$ (top); histogram $\histo{\subject_1^*}^{(3)}$ associated with the most central subject $\subject_1^* \in \subjects^{(3)}_1$ (alive, TTLFU $938$ days) of the first sub-population $\subjects^{(3)}_1$, obtained by feeding $\distances^{(3)}$ to the AHC clustering algorithm (left); and corresponding histogram $\histo{\subject_2^*}^{(3)}$ of the central subject $\subject_2^* \in \subjects^{(3)}_2$ (alive, TTLFU $1523$ days) of the other sub-population $\subjects^{(3)}_2$ (right).}
    \label{fig:histo_f3}
\end{figure*}

\begin{figure*}[!t]
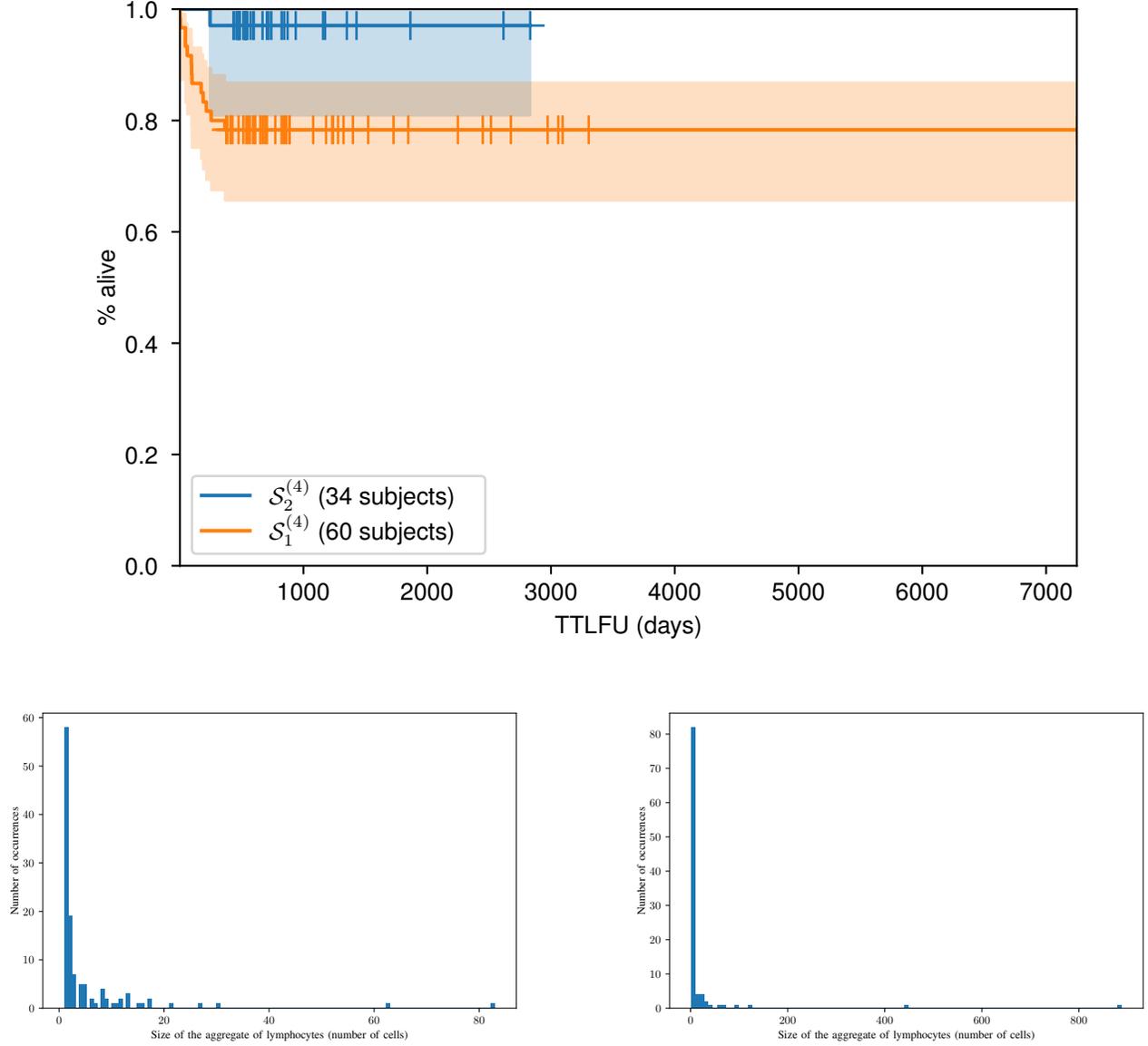

    \centering
    \input{survival_f4.pgf} \\
    \scalebox{0.47}{\input{f4_cluster1.pgf}} \hfill \scalebox{0.47}{\input{f4_cluster2.pgf}}
    \caption{Survival curves associated with both sub-populations $\subjects^{(4)}_1$ and $\subjects^{(4)}_2$ obtained by AHC using $\distances^{(4)}$ (top); histogram $\histo{\subject_1^*}^{(4)}$ associated with the most central subject $\subject_1^* \in \subjects^{(4)}_1$ (alive, TTLFU $418$ days) of the first sub-population $\subjects^{(4)}_1$, obtained by feeding $\distances^{(4)}$ to the AHC clustering algorithm (left); and corresponding histogram $\histo{\subject_2^*}^{(4)}$ of the central subject $\subject_2^* \in \subjects^{(4)}_2$ (alive, TTLFU $487$ days) of the other sub-population $\subjects^{(4)}_2$ (right).}
    \label{fig:histo_f4}
\end{figure*}

\begin{figure*}[!t]
    \centering
    \input{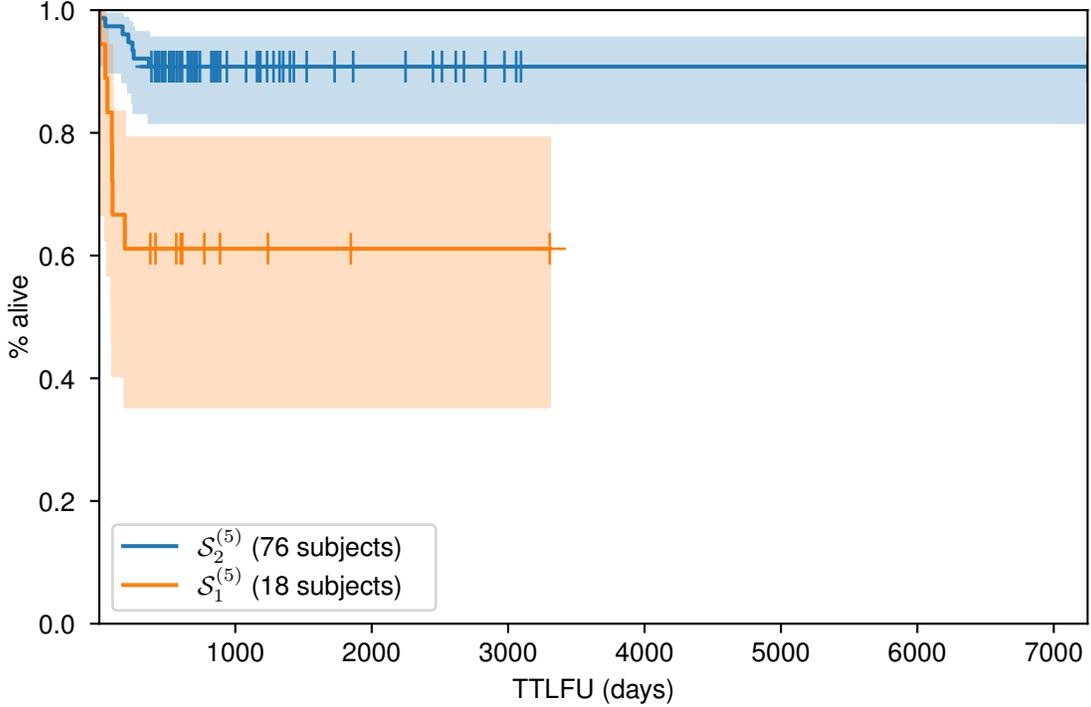} \\
    \scalebox{0.47}{\input{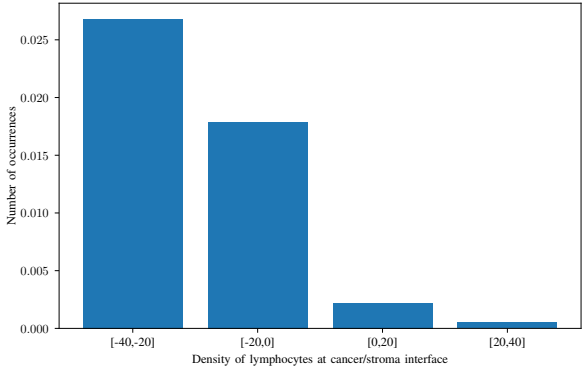}} \hfill \scalebox{0.47}{\input{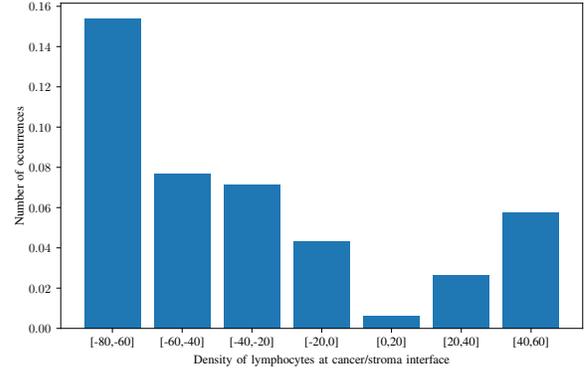}}
    \caption{Survival curves associated with both sub-populations $\subjects^{(5)}_1$ and $\subjects^{(5)}_2$ obtained by AHC using $\distances^{(5)}$ (top); histogram $\histo{\subject_1^*}^{(5)}$ associated with the most central subject $\subject_1^* \in \subjects^{(5)}_1$ (alive, TTLFU $773$ days) of the first sub-population $\subjects^{(5)}_1$, obtained by feeding $\distances^{(5)}$ to the AHC clustering algorithm (left); and corresponding histogram $\histo{\subject_2^*}^{(5)}$ of the central subject $\subject_2^* \in \subjects^{(5)}_2$ (dead, TTLFU $47$ days) of the other sub-population $\subjects^{(5)}_2$ (right).}
    \label{fig:histo_f5}
\end{figure*}

We show in Figures \ref{fig:histo_f1}, \ref{fig:histo_f2}, \ref{fig:histo_f3}, \ref{fig:histo_f4} and \ref{fig:histo_f5} (top) the survival curves obtained by AHC for each feature $\feature^{(i)} \in \features$, for which we are reporting results in Table \ref{tab:results}. In these figures, the orange curve is the survival curve for the cluster of subjects that are homogeneous to the feature ($\subjects^{(i)}_1$), and the blue curve is the survival curve for the cluster of heterogeneous subjects ($\subjects^{(i)}_2$). Shaded areas delimit a confidence interval of $95\%$.

Additionally, we show in these figures two histograms per feature. The first one (left) is the most central histogram within the cluster of homogeneous subjects, \emph{i.e.}, the histogram $\histo{\subject^*}^{(i)}$ that minimizes:
\begin{equation}
    \histo{\subject^*}^{(i)} = \arg\min\limits_{\subject \in \subjects^{(i)}_j} \sum\limits_{\subject' \in \subjects^{(i)}_j} \distanceH\left( \histo{\subject}^{(i)}, \histo{\subject'}^{(i)} \right) \;,
\end{equation}
where $j \in \{1, 2\}$ is the considered sub-population. The second histogram (right) is the most central within the heterogeneous cluster, using that same centrality measure.

The corresponding subjects can be seen as representatives for their clusters with respect to the feature $\feature^{(i)} \in \features$ used. Analyzing their histograms allows one to have an easy interpretation of the results of the clustering algorithm, which helps the pathologist understand the key aspects within distributions.

\subsection{Results obtained by combining features}
\label{sec:combination_results}

Now that we have studied the proposed features individually, we propose to merge those that have shown enough significance in survival prognosis into a unique matrix $\distances$, as presented in Section \ref{sec:combination}.

In order to fix the values of weights $\{ \weight^{(i)} \}_i$ in Equation \ref{eq:global}, we propose to give more importance to features that are more statistically significant than the others. For a feature $\feature^{(i)} \in \features$, let $p^{(i)}$ be the $p$-value of the log-rank test between survival curves of clusters $\subjects^{(i)}_1$ and $\subjects^{(i)}_2$, as reported in Table \ref{tab:results}. We choose $\weight^{(1)} = 0$ -- due to low significance of the feature -- and $\{ \weight^{(i)} \}_{i \in \{2, \dots, |\features|\}}$ as follows:
\begin{equation}
    \weight^{(i)} = \log \left( \frac{1}{p^{(i)}} \right) \;.
    \label{eq:weights}
\end{equation}

The $\log$ in Equation \ref{eq:weights} has been introduced to prevent very significant features such as $\features^{(2)}$ and $\features^{(5)}$ to absorb the contribution of other features.

Using the matrix $\distances$ computed as in Equation \ref{eq:global} with weights in Equation \ref{eq:weights}, we then use AHC and obtain two subpopulations $\subjects_1$ and $\subjects_2$, of which Kaplan-Meier survival curves are given in Figure \ref{fig:global}. The log-rank test between these curves has a $p$-value of $7.36 \cdot 10^{-8}$, which indicates a very strong significance of the clustering.

\begin{figure*}[!t]
    \centering
    \input{survival_voting.pgf}
    \caption{Survival curves associated with both sub-populations $\subjects_1$ and $\subjects_2$ obtained by AHC using $\distances$.}
    \label{fig:global}
\end{figure*}

\subsection{Discussion}

A first observation in our results is that Feature $\feature^{(1)}$ is not significant enough ($p\text{-value} > 0.05$). As a reminder, this feature captures the distances between lymphocytes and their closest cancer cells. When introducing it, we expected to find that subjects with a good survival prognosis would tend to have a strong proximity between such cells, as having lymphocytes in contact with cancer cells would indicate an immune response to the tumor. However, a more careful inspection of the feature reveals that it does not take into account the information that some lymphocytes are infiltrated, while others are not. It follows that the algorithm cannot distinguish between subjects that have lymphocytes close to cancer cells deep in the tumor, from those that only have lymphocytes in the stroma environment at the interface with the tumor, without any sort of infiltration (in both cases, corresponding histograms would show a peak in low positive bins).

This notion of infiltration of lymphocytes within the tumor has however been introduced into features $\feature^{(2)}$ and $\feature^{(5)}$, which correlate with survival with high confidence. The former feature captures the distance between lymphocytes and the tumor/stroma interface, associating a negative distance when lymphocytes are located in the tumor environment. The latter can be understood as a refinement of the former, and measures the density of lymphocytes among cells as distance to that interface increases. Again, a negative distance distinguishes bands that are within the tumor environment from those that are within the stroma environment.

When looking in more details at central histograms in Figure \ref{fig:histo_f2} (bottom), it reveals that both subjects seem to have both infiltrating and non-infiltrating lymphocytes. The absence of bins around distance $0\mu m$ is an artifact of the method used to determine the cancer/stroma interface. By construction, a triangle in the Delaunay triangulation of cells is part of the interface if it links at least one cancer and one stroma cell (see Section \ref{sec:feature_2}). As a consequence, lymphocytes are not used to define the border, which leads to this artifact. A noticeable difference between both histograms is that the central subject in the homogeneous sub-population $\subjects^{(2)}_1$ has a less deep lymphocyte infiltration (around $-40\mu m$) than the central subject in the heterogeneous sub-population $\subjects^{(2)}_2$ (around $-80\mu m$). When looking at the corresponding survival curves in Figure \ref{fig:histo_f2} (top), it reveals that subjects in the cluster $\subjects^{(2)}_2$ have a better survival prognosis than the other ones, which matches known facts regarding the link between TILs and survival outcome \cite{Saltz2018}. In order to verify if that observation generalizes to all subjects within the clusters -- and not only to the central subjects --, we have listed the minimum distance per histogram and have analyzed the distributions of such values per cluster. We obtain for cluster $\subjects^{(2)}_1$ a mean of $-43.29\mu m$, with standard deviation $20.56\mu m$; and for cluster $\subjects^{(2)}_2$ a mean of $-112.66\mu m$, with standard deviation $75.60\mu m$. Mann-Whitney U test \cite{mann1947test} indicates that both distributions are statistically significant ($U = 201$, $p\text{-value} = 1.85e-09$), which validates the observation.

A similar observation can be done with feature $\feature^{(5)}$. Histograms in Figure \ref{fig:histo_f5} show that the cental subject of the homogeneous cluster $\subjects^{(5)}_1$, which has a poorer survival prognosis, has a less deep infiltration of lymphocytes (around $-30\mu m$) when compared with the central subject of the heterogeneous cluster $\subjects^{(5)}_2$ (around $-70\mu m$). As for feature $\feature^{(2)}$, we have listed minimum band per histogram and have analyzed the distributions of such values per cluster. We obtain for cluster $\subjects^{(5)}_1$ a mean of $-27\mu m$, with standard deviation $5.6\mu m$; and for cluster $\subjects^{(2)}_2$ a mean of $-87\mu m$, with standard deviation $53.6\mu m$. Mann-Whitney U test indicates that both distributions are statistically significant ($U = 7.5$, $p\text{-value} = 3.56e-11$), which confirms what has been discussed in the previous paragraph.

Features $\feature^{(3)}$ and $\feature^{(4)}$ consider a coarser scale in the tissue, and study the link between presence of aggregates of lymphocytes at the tumor periphery and overall survival. The former feature captures the distances between aggregates, while the latter simply studies the distribution of the sizes of these aggregates. It can be seen in Figure \ref{fig:histo_f3} that the central subject in the homogeneous cluster $\subjects^{(3)}_1$ has a maximum distance between aggregates of lymphocytes (around $180\mu m$) which is much lower than for the central subject of the heterogeneous cluster $\subjects^{(3)}_2$ (around $390\mu m$). Survival curves show that the heterogeneous sub-population has a poorer survival prognosis. A closer analysis at histograms reveals that subjects from the homogeneous cluster have aggregates of lymphocytes that are in general close to one another, suggesting a uniform repartition of aggregates of lymphocytes to fight tumor progression, which denotes an efficient immune response. We have listed the maximum inter-aggregate distance per histogram and have analyzed the distributions of such values per cluster. We obtain for cluster $\subjects^{(3)}_1$ a mean of $309.63\mu m$, with standard deviation $213.06\mu m$; and for cluster $\subjects^{(3)}_2$ a mean of $431.16\mu m$, with standard deviation $274.72\mu m$. Mann-Whitney U test indicates that both distributions are statistically significant ($U = 611$, $p\text{-value} = 0.008$), which validates the observation.

In addition to this analysis of repartition of aggregates of lymphocytes, Figure \ref{fig:histo_f4} shows us that the size of these aggregates is important for survival prognosis. As a matter of fact, it can be seen that the most central subject of the homogeneous cluster $\subjects^{(4)}_1$ has aggregates that are a lot smaller (maximum around $86\mu m$) than those of the central subject of the heterogeneous cluster $\subjects^{(4)}_2$ (maximum around $850\mu m$). Since subjects in the homogeneous cluster have a poorer survival prognosis, this suggests that presence of large aggregates of lymphocytes may be characteristic of an efficient immune response. We have listed the maximum aggregate size per histogram and have analyzed the distributions of such values per cluster. We obtain for cluster $\subjects^{(4)}_1$ a mean of $300.35$ lymphocytes in the largest aggregate, with standard deviation $491.56$; and for cluster $\subjects^{(4)}_2$ a mean of $1773.97$ lymphocytes in the largest aggregate, with standard deviation $1579.39$. Mann-Whitney U test indicates that both distributions are statistically significant ($U = 192.5$, $p\text{-value} = 7.17e-10$), which validates the observation.

Finally, Figure \ref{fig:global} shows that the matrix of weighted co-occurrences of subjects within clusters significantly separates the two survival curves. This suggests that sub-populations produced tend to be coherent across features, especially for subjects that die after a very short period. As a consequence, this aggregation method offers an interesting approach to estimate survival prognosis based on a multi-criteria analysis.

\section{Conclusion}

In this article, we have proposed a methodology that allows comparing and clustering subjects on the basis of complex attributes, which capture their repartitions of cells within biopsies. To motivate our approach, we have introduced five features, each capturing some information on spatial repartitions of cells, and have shown that our method could produce clusters that are significantly different, while offering some interpretability of the results. We have illustrated our method on a dataset of H\&E biopsies of subjects suffering from Stage I lung adenocarcinoma, and have been able to find clusters that match existing knowledge in survival prognosis estimation.

Our method therefore offers a novel way of integrating complex features in survival analysis, in an automatic way. This allows one to explore new hypotheses regarding the links between cell organization patterns and survival outcome, and offers a systemic method for comparing two subjects on the basis of their personal attributes, which can help suggest someone a treatment that was efficient to a similar subject. Additionally, our method provides pathologists with very readable histograms of complex organizations of cells (or coarser elements such as cell aggregates) within tissues, which can be very difficult to assess by human eye. Our approach can thus help them produce a more detailed and quicker diagnosis, thus reducing the time between diagnosis and treatment of the cancer.

Perspective for extending this work are numerous. One direction consists in extending our catalogue of features to discover new distributions that may correlate with prognosis. Another direction is to evaluate our methodology on more complex data, such as mIHC, and other families of cancers. Finally, we want to explore variations of the elements in our approach, such as using different clustering algorithms, or changing the number of sub-populations produced, in case we manipulate a database of subjects with more than two profiles.


%



\section*{Acknowledgment}

The authors would like to thank Prof. Pascal Frossard and all members of the LTS4 laboratory at EPFL, Lausanne, Switzerland; as well as the group of Dr. Michel Cuendet at CHUV, Lausanne, Switzerland, for the interesting discussions which eventually led to that work.

\ifCLASSOPTIONcaptionsoff
  \newpage
\fi



\bibliographystyle{IEEEtran}
\bibliography{IEEEabrv,bibliography}
%

%








\end{document}